\documentclass[12pt,a4paper]{article}

\usepackage{amsmath}
\usepackage{graphicx}
\usepackage{times}
\usepackage{cite}
\usepackage{ulem}
\begin{document}
\begin{titlepage}
\title{On the real part  of  elastic scattering amplitude}
\author{ S.M. Troshin, N.E. Tyurin\\[1ex]
\small  \it NRC ``Kurchatov Institute''--IHEP\\
\small  \it Protvino, 142281, Russian Federation,\\
\small Sergey.Troshin@ihep.ru
}
\normalsize
\date{}
\maketitle

\begin{abstract}
	We discuss dominance of imaginary part of the elastic scattering amplitude and argue in favor  of  approximation  based on this dominance.
\end{abstract}
\end{titlepage}
\setcounter{page}{2}
\section*{Introduction}

The  measurements at the LHC energy of $\sqrt{s}=13$ TeV  have indicated \cite{tamas} that the hadron  interaction region responsible for the inelastic processes is transforming from a black disk   to a black ring  with  reflective area in its center.  Transition  to this  picture  has been discussed in \cite{trans} and references therein.
The black ring picture naturally emerges from the reflective scattering mode (antishadowing) \cite{plb93}.  Real part of the scattering amplitude is usually neglected under consideration of the shadow and reflective scattering  modes.
This is a common assumption based on the small value of  real part  compared  to the imaginary part of the scattering amplitude. 

On the other hand, the knowledge  of the real part  is important for testing validity of the dispersion relations  and are used for  conclusions on the  energy dependence of the total cross-section  \cite{bronz,mwu}. Real part is also sensitive  to possible nonlocality of the interaction and existence of extra dimensions. Reviews of the  issues related to the real part role   can be found in \cite{khuri, kang}. The nowadays interest to the real part   is motivated in particular by  odderon studies \cite{odd}. 

In this note, we argue in favor of the imaginary part dominance of the elastic scattering amplitude.
\section{Unitarity and real part of elastic scattering amplitude}
The  maximal odderon  and the Froissaron  contributions provide similar behavior for real and imaginary parts of the forward scattering amplitude  at $s\to\infty$:
\begin{equation}\label{max}
\mbox{Im} F(s,t=0)\sim s\ln^2 s,\,\,\,
\mbox{Re} F(s,t=0)\sim s\ln^2 s.
\end{equation}
Such a behavior of the scattering amplitude is inconsistent with  saturation of unitarity \cite{me}. 
   
 Eq. (\ref{max}) implies that the ratio 
\begin{equation}\label{ro}
\rho(s)\equiv \mbox{Re} F(s,t=0)/
\mbox{Im} F(s,t=0)\to const\neq 0.
\end{equation} at $s\to\infty$. The  model parametrization \cite{ter} gives an asymptotic value of $\rho=-0.2$\footnote{Respective value for $\bar p p$--scattering is $+0.2$.}in $pp$-scattering predicting  sign changing since  the experimental data at  highest available energies  hold positive values.  

It would be fair to note that consistency of the maximal odderon approach with the standard  constraints of quantum field theory  has been discussed  \cite{gar} in the frame of eikonal form of the elastic scattering amplitude  and with the {\it bounded phase} in the limit $s\to\infty$. The {\it ad hoc} $\theta$--like dependence  of the phase on the {\it ratio} $b/\ln s$ ($b$ is the   impact parameter) has been assumed. 

Meanwhile, unitarity saturation suggests the  positive asymptotic energy dependence of $\rho(s)$ decreasing like $1/\ln s$ \cite{usat}:
\begin{equation}\label{usat}
\rho (s)\sim \sigma_{inel}(s)/\sigma_{tot}(s).
\end{equation}
Saturation of the black disc limit  ($\sigma_{el}(s)/\sigma_{tot}(s)= \sigma_{inel}(s)/\sigma_{tot}(s)=1/2$) assumes then energy independent {\it nonzero} limit of $\rho(s)$ at $s\to\infty$.

For  discussions of the real part of the elastic scattering amplitude based on  unitarity, it is convenient to consider the elastic scattering amplitude $f(s,b)$ in the impact parameter representation \cite{webb}. The unitarity relation for  
$f(s,b)$ can be written in the form:
\begin{equation}\label{unit}
[\mbox{Re}f(s,b)]^2=\mbox{Im}f(s,b)[1-\mbox{Im}f(s,b)]-h_{inel}(s,b),
\end{equation}
where $h_{inel}(s,b)$ is the inelastic overlap function, which has  nonnegative values.  Then, it is evident that
\begin{equation}\label{uin}
h_{inel}(s,b)\leq \mbox{Im}f(s,b)[1-\mbox{Im}f(s,b)]
\end{equation}
and
\begin{equation}\label{unin}
[\mbox{Re}f(s,b)]^2\leq \mbox{Im}f(s,b)[1-\mbox{Im}f(s,b)].
\end{equation}
When $\mbox{Im}f \to1$, the real part $\mbox{Re}f\to 0$ at $s\to\infty$ implying the  above noted inconsistency of maximal odderon with the unitarity saturation\footnote{Inconsistency of the maximal odderon with the black disk limit saturation has been demonstrated in \cite{fin}.}.

 The analysis of the  experimental data at  highest available energy $\sqrt{s}=13 $ TeV
\cite{tamas} resulted in the following approximate relation:
\begin{equation}\label{bd}
[\mbox{Re}f(s,b)]^2+[\mbox{Im}f(s,b)-1/2]^2\simeq 0,
\end{equation}
valid in the impact parameter range $ 0\leq b \leq 0.4$ fm.  
It describes behavior of the elastic scattering amplitude at the LHC energy  $\sqrt{s}=13 $ TeV and summarizes   the experimental situation  in the  impact parameters' range  of maximal importance. This Eq. (\ref{bd}) implies  that $\mbox{Re}f(s,b)\simeq 0$, $\mbox{Im}f(s,b)\simeq 1/2$ with good accuracy. It  demonstrates that {\it  small value of the  ratio $\rho$, Eq. (\ref{ro}), at this energy value  is probably not due to the maximal odderon contribution} since   the real part contribution  is small  in the central region of the impact parameters whereas imaginary part of the scattering amplitude dominates. Similar  results for  the real part  contribution have been obtained at lower energies in \cite{alkin}.  

The Eq. (\ref{unit}) can be rewritten as an expression for the inelastic overlap function $h_{inel}(s,b)$ indicating no dependence of this function (and respectively the inelastic cross--section $\sigma_{inel}(s)$) from  the  sign of the real part of the elastic scattering amplitude.  

The real part  contribution does not play also a   role at large impact parameter values since in this $b$ range: 
\begin{equation}
h_{inel}(s,b)/\mbox{Im}f(s,b)\to 1,
\end{equation}
and
\begin{equation}
[\mbox{Re}f(s,b)]^2<<h_{inel}(s,b)\,\,\mbox{and}\,\,\mbox{Im}f(s,b)<< 1
\end{equation}
due to  Gribov--Froissart projection formula \cite{coll}:
\begin{equation}\label{grf}
f(s,b)\simeq \omega (s)\exp{(-\mu b)}.
\end{equation}
According to Eq. (\ref{grf}), the amplitude $f(s.b)$ should be represented by a factorized function of  $s$ and $b$ variables at large values of $b$.
A hadron can be seen as  a structure with  hard core coated with a fragile layer \cite{cent}.
We do not consider here possible effects of  real part  related to New Physics \cite{khuri}. 

Thus, there is no room for any significant real part contribution,
and the elastic scattering amplitude being a predominantly imaginary justifies the use of standard approximation by a pure imaginary function.  Quantitative smallness of a real part  contribution should not accompanied by any qualitative speculations based the real part accounting expanded  with  {\it ad hoc} various assumptions on the respective energy dependence. 

The unitarity solution  can be given in the form of the generalized reaction matrix  approach \cite{ech}:
\begin{equation}\label{um}
F=F[U]
\end{equation}
 where  the function $U$ is used as an input and  object for building models.
Equation (\ref{um}) represents the relativistic generalization of the basic equation of the quantum theory of radiation damping \cite{hit}, which implies {\it depletion of the transient state}and provides solution for the partial amplitude in the rational  form:
\begin{equation}\label{f}
f(s,b)=u(s,b)/[1-iu(s,b)],
\end{equation}
where the complex input function $u(s,b)$  is a subject for a model constructions [with an evident constraint $\mbox{Im}u(s,b)\geq 0$]  \cite{prd}.
Then the discussed 
 approximation for the scattering amplitude based on the dominance of its imaginary part is translating itself to relative  vanishing of real part   of the input function $U(s,t)$   
 connected with  $u(s,b)$ by  Fourier-Bessel transformation. 

Dominance of its imaginary part corresponds to a leading role of the {\it inelastic} intermediate states  contributions. Elastic intermediate states do not contribute to the imaginary part of this function.

Eq. (\ref{f}) covers {\it the full unitary circle} \cite{map}.  Analytical properties of the function $u(s,\beta)$  ($\beta = b^2$) are determined by the relation
\begin{equation}\label{beta}
u(s,\beta)=\frac{\pi ^2}{s}\int_{t_0}^\infty \tilde \rho (s,t')K_0(\sqrt { t'\beta})dt',
\end{equation} 
where $\tilde\rho$ denotes the spectral density  of representation:
\begin{equation}\label{spec}
U(s,t)=\int_{t_0}^\infty \frac{\tilde \rho (s,t')}{t'-t}dt'.
\end{equation}
Here $t_0$ is  position of the lowest $t$--channel singularity   and $K_0$ is Macdonald function. 
Eq. (\ref{beta}) can be used to obtain the explicit form of  function $u(s,\beta)$ and its real and imaginary parts are determined by the respective parts of the spectral density $\tilde \rho(s,t)$. 

Eq. (\ref{beta})  leads to the factorized form of the function $u(s,\beta)$ with product of the energy dependent {\it complex function} and another  impact parameter dependent real function exponentially decreasing at large values of the impact parameter\footnote{The function $K_0(z)$ v at large values of $z$ is approximated as $K_0(z)\simeq \sqrt{\frac{\pi}{2z}}\exp(-z)$.}. Such  form is similar to Eq. (\ref{grf}). One should note that factorization of the function $u(s,b)$  does not lead to factorization of the scattering amplitude $f(s,b)$ in the whole region of variation of $b$, only in the region of large values of $b$. 

The  expressions for the  inelastic overlap function $h_{inel}(s,b)$ and the real part  
$\mbox{Re}f(s, b)$ of elastic scattering amplitude are:
\begin{equation}\label{i}
h_{inel}(s,b)=\mbox{Im}u(s,b)/|1-iu(s,b)|^2,
\end{equation}
\begin{equation}\label{ii}
\mbox{Re}f(s,b)=\mbox{Re}u(s,b)/|1-iu(s,b)|^2.
\end{equation}
Evidently, Eqs. (\ref{f}) and (\ref{i}), (\ref{ii})  imply  the different  dependencies of $\sigma_{tot}(s)$ and $\sigma_{el}(s)$ compared to the dependencies of 
$\sigma_{inel}(s)$ and $\mbox{Re}F(s,t=0)/s$ at $s\to\infty$. 

 Eq. (\ref{usat}) can   be derived assuming  the ratio  $\mbox{Re}u(s,b)/\mbox{Im}u(s,b)$ is a small positive constant. This constant represents a   proportionality factor in Eq. (\ref{usat}) and also in relation of $\mbox {Re} F(s,t)$ with the inelastic overlap function ${H_{inel}}(s,t)$. It should be noted that  positivity of this constant is consistent with positivity of the real part of the forward scattering amplitude
 \cite{mwu}
 and is correlated with  positivity of the imaginary part of the scattering amplitude. 

Eqs. (\ref{f}) and   (\ref{ii}) imply also a difference in $t$--dependencies of the functions  
$\mbox{Im}F(s,t)$ and  $\mbox{Re}F(s,t)$. It  results, in particular, in a complicated structure of the differential cross--section  in the  diffraction cone region. The noted difference appears as a consequence of the amplitude unitarization.

\section*{Conclusion}
Finally, we would like to note that  the  scattering picture that follows from  dominance of  the imaginary part of elastic  amplitude over its real part seems to be in a full compliance  with results of the current experimental and theoretical studies. The impact parameter analysis of the LHC data implies transition from a  black disc to a black ring picture of hadron interactions. This conclusion  remains to be valid  and leaves almost no room for the other interpretations.


\begin{thebibliography}{99}
\bibitem{tamas}
T.  Cs\"{o}rg\H{o}, R. Pasechnik, and A. Ster,  Acta Phys. Pol. B Proc. Suppl. {\bf 12 }, 779  (2019).
\bibitem{trans}
S.M. Troshin and N.E.  Tyurin, { Phys. Lett. B} {\bf 855}, 138871  (2024). 
\bibitem{plb93}
S.M. Troshin and N.E.  Tyurin, { Phys. Lett. B} {\bf 316}, 175  (1993).
\bibitem{bronz}
J.B. Bronzan, G.L. Kane, and U.P. Sukhatme, { Phys. Lett. } {\bf 49B}, 272 (1974).
\bibitem{mwu}
A. Martin, T.T. Wu, {Phys. Rev. D}{\bf 97}, 014011 (2018).
\bibitem{khuri}
N.N. Khuri, in Proc. of Vth Blois Workshop ---Int. Conf. Elastic and Diffractive Scattering, Providence, RI, June 8--12, 1993, eds. H.M. Fried, K. Kang, and C.-I- Tan (World Scientific, 1994), p.42.
\bibitem{kang}
K. Kang, P. Valin, and A.R. White, in Proc. of Vth Blois Workshop ---Int. Conf. Elastic and Diffractive Scattering, Providence, RI, June 8--12, 1993, eds. H.M. Fried, K. Kang, and C.-I- Tan (World Scientific, 1994), p.50.
\bibitem{odd}
E. Martynov and B. Nicolescu, EPJ Web Conf. {\bf 206}, 06001 (2019) .
\bibitem{me}
S.M. Troshin,  { Phys. Lett. B} {\bf 682}, 40  (2009).
\bibitem{ter}
E. Martynov and G. Tersimonov, Phys. Rev. D {\bf 100}, 114039 (2019).
\bibitem{gar}
P. Gauron, L. \L ukaszuk, and B. Nicolescu, Phys. Lett. B {\bf 294}, 298 (1992).
\bibitem{usat}
S.M. Troshin and N.E.  Tyurin, {Mod. Phys. Lett. A} {\bf 33}, 150206  (2018). 
\bibitem{webb}
B.R. Webber, Nucl. Phys. B {\bf 87}, 269 (1975).
\bibitem{fin}
J. Finkelstein, H.M. Fried, K. Kang, and C-I. Tan, Phys. Lett. B {\bf 232}, 257 (1989).
\bibitem{alkin}
A. Alkin, E. Martynov, O. Kovalenko, and S.M. Troshin,
Phys. Rev. D {\bf 89},  091501 (2014).
\bibitem{coll}
P.D.B. Collins, An Introduction to Regge Theory and High Energy Physics, Cambridge University Press: Cambridge, UK, 1977; 460p.
\bibitem{cent}
S.M. Troshin and N.E.  Tyurin, { Phys. Lett. B} {\bf 816}, 136186  (2021).
\bibitem{ech}
V.I. Savrin, N.E. Tyurin, and O.A. Khrustalev,  Sov. Journ. of Elementary Particles and Atomic Nuclei {\bf 7}, 93 (1976).
\bibitem{hit}
W. Heitler, Quantum theory of radiation, Oxford University Press: Oxford, UK, 1954.
\bibitem{prd}
S.M. Troshin and N.E.  Tyurin, { Phys. Rev. D} {\bf 49}, 4427  (1994).
\bibitem{map}
S.M. Troshin and N.E.  Tyurin, {Mod. Phys. Lett. A} {\bf 38}, 235061  (2023).

\end{thebibliography}
\end{document}